\RequirePackage{fix-cm}

\documentclass[smallextended]{svjour3}       
\smartqed 

\usepackage{pgfplots} 
\pgfplotsset{compat=1.7}
\usepackage[utf8]{inputenc}
\usepackage{authblk}
\usepackage{amssymb}
\usepackage{amsmath}
\usepackage{amsfonts}
\usepackage{mathrsfs}
\usepackage{pxfonts}
\usepackage{latexsym}
\usepackage{graphicx}
\usepackage{hyperref}
\usepackage{float}
\usepackage{epstopdf,setspace}
\begin{document}

\title{The scalar field and gravity combined variable in ADM theory}

\author{Avadhut V Purohit}

\institute{ \at
              Chennai Mathematical Institute, Kelambakkam, India -603103 \\
              \email{avdhoot.purohit@gmail.com}           
}

\date{}

\maketitle

\begin{abstract}
The canonical theory of gravity together with the scalar field is written as a combined 
variable theory in ADM form, in the classical and quantum theory. 
FLRW $\kappa = 0$ cosmology is rewritten in the combined variable theory.
\end{abstract}

\section{Introduction}
\hspace*{0.5cm}ADM theory of space-time and gravity is extended to include the scalar 
field. A combined variable representation is defined in both classical and quantum theory. 
Equations of field theory and Hamiltonian are derived. The role of a lapse 
function and shift vector in the combined variable theory is studied. Limits of no 
gravity and no scalar field are found. Quanta of the combined variable are obtained. 
In FLRW $\kappa = 0$ metric of cosmology, the combined variable theory gives the 
evolution of space-time. Results about the energy spectrum and singularity are 
obtained. \newline
\hspace*{0.5cm} Although the combined variable theory is built on
the ADM formulation, unlike Wheeler-DeWitt theory [\ref{item4}], $q_{ab}$ and $\phi$ remain parameters and combined variables $\Phi$ and its conjugate momentum are observables. There have been attempts [\ref{item9}-\ref{item10}] to quantize similar field in the cosmological context using path integral formulation. Such field in these papers represents a baby Universe. Where as, quantum of combined variable is a quantum of both gravity and scalar field together.
The combined variable field have geometric energy spectrum. Gravity is quantized independent of matter fields in Loop Quantum Gravity (refer [\ref{item2}],[\ref{item5}]). But the combined variable theory suggests that gravity cannot be quantized without matter field.\newline
\hspace*{0.5cm}
Classical theory is developed in the first section. Time-dependent and space-dependent 
parts of both fields are separated. Then, the full Hamiltonian constraint equation 
is re-interpreted as a classical field equation for the combined variable field. 
Hamiltonian of the combined variables allow to quantize the theory in a standard 
way. This is done in the third section. In section four combined variable ADM theory is 
applied to FLRW $\kappa = 0$ cosmology and section five concludes earlier 
three sections. Throughout the paper $16\pi G=1$, $\hbar = 1$ and $c=1$ is set. 
All calculations are done in MATHEMATICA.  

\section{Classical Theory}
The action for a scalar field in presence of ADM gravity is given as  
\begin{equation} \label{eqn1}
 \mathscr{A} =  \int_{\mathbb{R}}dt \int_{M} d^3 x \left( |N| \sqrt{q} \left\lbrace  \left( R^{(3)}+
  K^{ab}K_{ab} - K^2 \right) + 
   \left( \frac{1}{2} \dot{\phi}^{2} - \frac{1}{2} |\nabla \phi|^{2} - V(\phi) \right) \right\rbrace  \right) (t,\vec{x}) 
 \end{equation}
 ADM formulation \ref{item1} of general relativity developed by Richard Arnowitt, Stanley Deser, and Charles W. Misner is a canonical 
 formulation which is being studied over more than five decades. Refer (\ref{item2}
 , \ref{item3}) for detail analysis of ADM formulation. 
 In ADM formulation, 4-dimensional spacetime manifold is foliated into a family of 
 spacelike hypersurfaces. Shift vector together with lapse function decides the foliation of spacetime manifold. Canonical conjugate
 momenta corresponding to $q_{ab}$, shift vector $N^{a}$, lapse function $N$, and scalar field $\phi$ are respectively given
 (derived in  \ref{item3}, section 1.2, equation 1.2.1) as
  \begin{align}\label{eqn2}
 & P^{ab}(t,\vec{x}) \coloneqq \left( \frac{|N|}{N} \sqrt{q} \left( K^{ab} - K q^{ab} \right) \right) 
 (t,\vec{x}) \\
  & C_{a}(t,\vec{x}) \coloneqq \Pi_{a}(t,\vec{x}) = 0 \\
  & C(t,\vec{x}) \coloneqq \Pi(t,\vec{x}) = 0 \\
  & P_{\phi}(t,\vec{x}) \coloneqq \left( |N| \sqrt{q} \dot{\phi} \right)(t,\vec{x}) 
 \end{align}
As mentioned in \ref{item1} momentum $ P^{ab}$ refers to motion in the time leading out of the original $t = constant$ surface. 
Extrinsic curvature $K_{ab}$ describes how the normal to the surface converge or 
diverge. Only non-trivial Poisson brackets are given (borrowed from  \ref{item2}, equation 1.2.1.9) as 
\begin{align} \label{eqn3}
 & \left\lbrace q_{ab}(t,\vec{x}),  P^{cd}(t,\vec{x}^{\prime}) \right\rbrace = 
 \delta^{c}_{(a} \delta^{d}_{b)}  \delta^{(3)}(\vec{x},\vec{x}^{\prime}) \\
 &  \left\lbrace  \phi(t,\vec{x}), P_{(\phi)}(t,\vec{x}^{\prime}) \right\rbrace = \delta^{(3)}(\vec{x},\vec{x}^{\prime})
\end{align} 
There are no equations to determine lapse function $N$ and shift vector $N^{a}$. Therefore action reduces to 
\begin{equation}\label{eqn4}
 \mathscr{A} = \int_{\mathbb{R}}dt  \int_{M} d^3x \left\lbrace P^{ab}\dot{q}_{ab} + P_{\phi}\dot{\phi} - 
 \left( N^{a}\mathscr{H}_{a}+|N| \mathscr{H}_{\text{scalar}} + |N|\mathscr{H}_{\phi} \right)\right\rbrace (t,\vec{x})
\end{equation}
Where
\begin{align}\label{eqn5}
 & \mathscr{H}_{\text{scalar}} \coloneqq \left( \frac{1}{\sqrt{q}} \left( q_{ac}q_{bd}-
 \frac{1}{2}q_{ab}q_{cd} \right) P^{ab}P^{cd} - \sqrt{q} R^{(3)} \right) (t,\vec{x}) \\
 &\mathscr{H}_{\phi} = \sqrt{q} \left( \frac{P^{2}_{\phi}}{2N^2 
\text{det}(q)} + \frac{1}{2}|\nabla\phi|^{2} + V(\phi) \right) (t,\vec{x}) \\
& \mathscr{H}_{a} \coloneqq \left( -2 q_{ac}D_{b}P^{bc} \right) (t,\vec{x})
\end{align}
The total Hamiltonian is a sum of Hamiltonian scalar constraints of gravity, scalar field Hamiltonian 
and Diffeomorphism constraints (also called vector constraints). Refer equation 
1.2.6 of  \ref{item3} or section I.2.1,  \ref{item2} for detail discussion.
\begin{align}\label{eqn6}
	H_{\text{total}} = & \int d^{3}x  \left( \frac{N}{\sqrt{q}} \left( q_{ac}q_{bd} -
	\frac{1}{2}q_{ab}q_{cd} \right) P^{ab}P^{cd} - N \sqrt{q} R^{(3)} \right) \\ \nonumber
	 + & \int d^{3}x \hspace{0.1 cm} N\sqrt{q} \left( \frac{P^{2}_{\phi}}{2N^2 
	\text{det}(q)} + \frac{1}{2}|\nabla\phi|^{2} + V(\phi) \right)  \\  \nonumber
	- 2 & \int d^{3}x \hspace{0.1 cm} N^{a}q_{ac}D_{b}P^{bc}
\end{align}
Separate gravitational field variables $q_{ab} \coloneqq q_{a}(t) \tilde{q}_{ab}(\vec{x})$ and 
$P^{ab} \coloneqq P^{a}(t) \tilde{P}^{ab}(\vec{x})$.
\begin{equation}\label{eqn7}
	- \left( \frac{q_{a}q_{b}P^{a}P^{b}}{\sqrt{\text{det }q_{a}}}\right) (t) 
 \int d^{3}x 	\left( N \tilde{F}_{abcd} \tilde{P}^{ab}\tilde{P}^{cd} \right)(\vec{x}) 
\end{equation}
with $- \tilde{F}_{abcd} = \tilde{G}_{abcd} \coloneqq \frac{1}{\sqrt{\tilde{q}_{ab}}}\left( \tilde{q}_{ac}\tilde{q}_{bd} 
- \frac{1}{2} \tilde{q}_{ab}\tilde{q}_{cd} \right) $. $\tilde{G}_{abcd}$  is a spatial part of inverse DeWitt metric. DeWitt metric has
signature $(- + + + + + )$ which reduces to overall negative signature. 
Therefore $\tilde{F}_{abcd}$ is defined to be negative of inverse DeWitt metric. 
Properties of DeWitt metric are discussed in detail in  \ref{item4}, appendix A. Choose lapse function as  
\begin{equation}\label{eq87}
 N(t, \vec{x}) \coloneqq N(t) \tilde{N}(\vec{x})
\end{equation}
Then the first term becomes
\begin{equation}\label{eqn9}
- \left( N F_{ab}P^{a}P^{b}\right) (t) \int d^{3}x 
\left( \tilde{N}(\vec{x}) \tilde{F}_{abcd} \tilde{P}^{ab}\tilde{P}^{cd} \right)(\vec{x})
\end{equation}
with $F_{ab} \coloneqq \frac{q_{a}q_{b}}{\sqrt{\text{det }q_{a}}} $. The second term in the gravitational part of Hamiltonian 
\begin{equation}\label{eqn10}
	- N\sqrt{\text{det }q_{a}}(t) \int d^{3}x \tilde{N}(\vec{x})  \sqrt{\tilde{q}_{ab}}(\vec{x}) \hspace*{0.1 cm} R^{(3)} (t, \vec{x})
\end{equation}
Separate scalar field variables into $\phi(t, \vec{x}) \coloneqq \phi(t) \tilde{\phi}(\vec{x})$ and 
$P_{\phi}(t, \vec{x}) \coloneqq P_{\phi}(t) \tilde{P}_{\phi}(\vec{x})$. The first term in the scalar field part of the Hamiltonian
becomes
\begin{equation}\label{eqn11}
	 \frac{P_{\phi}^{2}}{N\sqrt{\text{det }q_{a}}} (t) \int d^{3}x  \frac{\tilde{P}_{\phi}^{2}}{2 \tilde{N}(\vec{x}) \sqrt{\tilde{q}_{ab}}}
\end{equation}
The second term as well as massive coupling term can be absorbed into single term as
\begin{equation}\label{eqn12}
	\frac{1}{2}\mu^{2}(\vec{x}) N \sqrt{\text{det }q_{a}}(t) \phi^{2}(t) 
\end{equation}
with 
\begin{equation}\label{eqn13}
	\mu^{2}(\vec{x}) \coloneqq \int d^{3}x^{\prime} \tilde{N}(\vec{x}^{\prime})\sqrt{\tilde{q}_{ab}} 
	\left( \big| \nabla \tilde{\phi} \big|^{2} + m^{2}\tilde{\phi}^{2} \right) (\vec{x^{\prime}})
\end{equation}
Combining all terms in the scalar Hamiltonian constraints as well as scalar field Hamiltonian,
\begin{equation} \label{eqn14}
	H_{\phi} + H_{\text{scalar}} = \tilde{T}_{\phi} \frac{P_{\phi}^{2}}{2N\sqrt{\text{det }q_{a}}} 
	- \tilde{T}_{\text{grav}} \left( N F_{ab}P^{a}P^{b}\right) + N V_{\text{CF}} 
\end{equation}
With 
\begin{align}\label{eqn15}
	& \tilde{T}_{\phi} \coloneqq \int \frac{\tilde{P}_{\phi}^{2}}{  \tilde{N} \sqrt{\tilde{q}_{ab}}} \hspace{2cm}
	 \tilde{T}_{\text{grav}} \coloneqq  \int d^{3}x \left( \tilde{N}\tilde{F}_{abcd} \tilde{P}^{ab}\tilde{P}^{cd} \right)(\vec{x}) \\ 
	& V_{\text{CF}} \coloneqq - \sqrt{\text{det }q_{a}} \int d^{3}x \tilde{N}\sqrt{\tilde{q}_{ab}} \hspace*{0.1 cm} R^{(3)} + 
	\frac{1}{2}\mu^{2} \sqrt{\text{det }q_{a}} \phi^{2} +
	\sqrt{\text{det }q_{a}} \int d^{3}x \tilde{N}\sqrt{\tilde{q}_{ab}} \hspace*{0.1 cm} V(\phi)
\end{align}
$\tilde{T}_{\phi}$ can be thought of kinetic energy of the spatial part of scalar field, $\tilde{T}_{\text{grav}}$
is kinetic energy due to the spatial part of gravity and $V_{\text{CF}}$ is combined potential of both fields. \newline
Symmetrizing the second term in terms of $q_{a}$ and $P^{a}$ in (\ref{eqn14})
\begin{equation} \label{sym}
	H_{\phi} + H_{\text{scalar}} = \tilde{T}_{\phi} \frac{P_{\phi}^{2}}{2N\sqrt{\text{det }q_{a}}} 
	- \tilde{T}_{\text{grav}} P^{a}N F_{ab}P^{b}+ N V_{\text{CF}} 
\end{equation}
In classical ADM theory, there is no issue of ordering gravitational field variables.
But in case of combined variable theory, different symmetric combinations may have
different combined variable field theoretic extensions. Which may or may not be equivalent. 
Other symmetric combinations such as $q_{a}P^{a}P^{b}q_{b}$ or linear combination of both, 
these combinations may not have simple combined variable field theoric extensions. Such extension may not 
even be allowed. It will be addressed in the next work. In this paper, only (\ref{sym}) combination is chosen. \\
\hspace*{0.5cm} Now, $\hat{H}| \Phi \rangle = 0 $ is interpreted as classical field equation for 
classical combined variable field $\Phi$. This definition is unique upto chosen symmetric combination
of gravitational field variables. 
\begin{equation}\label{eqn16}
	 -\left( \frac{\tilde{T}_{\phi}}{2N \sqrt{\text{det } q_{a}}} \frac{\partial^{2}}{\partial \phi^{2}} - 
	\tilde{T}_{\text{grav}}\frac{\partial}{\partial q_{a}} N F_{ab}\frac{\partial}{\partial q_{b}} - N V_{\text{CF}} \right) \Phi (\phi, q_{c}) = 0 
\end{equation}
OR
\begin{equation} \nonumber
	\left( \partial^{\mu}\eta_{\mu \nu} \partial^{\nu} -  V_{\text{CF}} \right) \Phi (\phi, \vec{q}) = 0
\end{equation}
Where the metric for superspace $(\phi, \vec{q})$ is defined as
\begin{equation}\label{eqn18}
	\eta_{\mu \nu} \coloneqq \left( \frac{1}{N\sqrt{\text{det } q_{a}}}(t)\tilde{T}_{\phi}(\vec{x}), \hspace{0.1cm} -
	 \tilde{T}_{\text{grav}}(\vec{x}) N(t) F_{ab}(t) \right)
\end{equation}
and $\partial^{\mu} = \left( \frac{\partial}{\partial \phi} , \frac{\partial}{\partial q_{a}} \right)$.
This allows us to write an action for combined variable field as, 
\begin{equation}\label{eqn19}
	\mathscr{A} = \int d\phi \int d^{D}q \hspace{1mm} \frac{1}{2} 
	\left( \frac{\tilde{T}_{\phi}}{2N\sqrt{\text{det } q_{a}}} \left( \frac{\partial \Phi}{\partial \phi} \right)^{2}
	- 	\tilde{T}_{\text{grav}} N \left( \frac{\partial \Phi}{\partial q_{a}} \right)F_{ab} 
	\left( \frac{\partial \Phi}{\partial q_{b}} \right) 
	- N V_{\text{CF}}\Phi^{2} \right)
\end{equation}
Invariance of action under $\Phi \rightarrow \Phi + \delta \Phi$ gives
\begin{equation}\label{eqn20}
	\left( \frac{\tilde{T}_{\phi}}{2N\sqrt{\text{det } q_{a}}} \frac{\partial^{2}}{\partial \phi^{2}} -  \tilde{T}_{\text{grav}}
	\frac{\partial}{\partial q_{a}} \left( NF_{ab}\frac{\partial}{\partial q_{b}} \right) 
	-N V_{\text{CF}} \right) \Phi (\phi, q_{c}) = 0 
\end{equation}
$D$ in (\ref{eqn19}) represents number of independent components of 3-metric $q_{a}(t)$. For 
example in case of $q_{1}(t) = q_{2}(t) = q_{3}(t) = q$, $D = 1$. The action chosen is
not unique but the simplest form of action which satisfies above field equation. \\
\hspace*{0.5cm} Vector constraints
\begin{equation}\label{eqn21}
	H_{\text{vector}} = -2 \int d^{3}x \hspace{0.1 cm} N^{a}q_{ac}D_{b}P^{bc} 
	= -2 \left( q_{a}P^{a} \right)  \left( \int d^{3}x \hspace{0.1 cm} N^{a}
	\tilde{q}_{ac}D_{b}\tilde{P}^{bc} \right) \\
	= \alpha_{\text{Diff}}(\vec{x})q_{a}P^{a}
\end{equation}
translate into combined variable theory as
\begin{equation}
	H_{\text{vector}} \Phi = -i \alpha_{\text{Diff}}(\vec{x})q_{a}\frac{\partial \Phi}{\partial q_{a}}
\end{equation}
There are two possibilities
\[ 
	0= 
	\begin{cases}
		\alpha_{\text{Diff}}(\vec{x}) \\
		q_{a}\frac{\partial \Phi}{\partial q_{a}}
	\end{cases}
\]
$q_{a}\frac{\partial \Phi}{\partial q_{a}}$ is directional derivative of combined variable field along
$q_{a}$. Assuming that to be zero would also make the second term in (\ref{eqn19}) zero
because $F_{ab} \coloneqq \frac{q_{a}q_{b}}{\sqrt{\text{det }q_{a}}}$ . Therefore, this 
assumption is not valid. This implies  
\begin{equation}\label{eqn22}
	\alpha_{\text{Diff}}(\vec{x}) \coloneqq -2\int d^{3}x \hspace{0.1 cm} N^{a}
	\tilde{q}_{ac}D_{b}\tilde{P}^{bc} = 0
\end{equation}
That means diffeomorphism constraints do not play any role in the dynamics of 
the combined variable field. It puts restrictions on the spatial parts of gravitational 
field variables.  
Unlike other fields gravity being a dynamical theory of space-time itself, it cannot evolve with respect to external time.
It evolves with respect to matter field. Therefore, momentum conjugate to the combined variable field is 
\begin{equation}\label{eqn23}
	\Pi \coloneqq \frac{\partial \mathscr{L}}{\partial \left( \frac{\partial \Phi}{\partial \phi}\right)}
	=  \frac{\tilde{T}_{\phi}}{2N \sqrt{\text{det } q_{a}}} \left( \frac{\partial \Phi}{\partial \phi} \right)
\end{equation}
where $\mathscr{L}$ is the Lagrangian density for combined variable theory. 
Hamiltonian for combined variable theory is obtained using Legendre transformation,
\begin{equation}\label{eqn24}
	\mathbf{H} = \int d^{D}q \hspace{1mm} \frac{N}{2} 
	\left( \left( \frac{2 \sqrt{\text{det } q_{a}}}{\tilde{T}_{\phi}} \right) \Pi^{2}
	+ 	\tilde{T}_{\text{grav}}  \left( \frac{\partial \Phi}{\partial q_{a}} \right) F_{ab}
	\left( \frac{\partial \Phi}{\partial q_{b}} \right) 
	+ V_{\text{CF}}\Phi^{2} \right)
\end{equation}
Although $\mathbf{H}$ is called as Hamiltonian, it should rather be considered as an 
observable which gives $\phi$ evolution. This Hamiltonian is different for different lapse function.  
Therefore choosing a particular lapse function is equivalent of selecting a particular scalar field. 
Evolution of $\Pi$ can be solved using 
\begin{equation}\label{eqn25}
	\frac{\partial \Pi}{\partial \phi} = \lbrace \Pi, \mathbf{H} \rbrace_{\text{P.B.}}
\end{equation}
\hspace*{0.5cm} Invariance of an action of the combined variable field under infinitesimal change
in the scalar field as well as 3-metric gives us stress-energy tensor. 
The procedure for obtaining this tensor (can be found in section 1.5 of \ref{item7}.
Identify $x^{\mu}$ with $q^{\mu}$ and $\phi$ as $\Phi$) is quite generic.
\begin{equation}\label{eqn26}
	T^{\mu}_{\nu} \coloneqq \frac{\partial \mathscr{L}}{\partial \left(\partial_{\mu}	\Phi\right)} 
	\partial_{\nu}\Phi- \delta^{\mu}_{\nu}\mathscr{L}
\end{equation}
Here, $\mu, \nu = 0$ is scalar field dependent component and 
$\mu, \nu = 1, 2, 3$ are gravitational components. $T^{0}_{0}$ is identified as the 
energy density of the combined variable field and $-T^{i}_{j}$ ($i, j = 1, 2,3$) as
 pressure. \\
 \textbf{Flat space-time:}  \newline
 If we take flat space-time limit of a complex combined variable field,
	the Hamiltonian reduces to 
 \begin{equation}
	 \mathbf{H} =  \frac{1}{2} \left( \left( \frac{2}{\tilde{T}_{\phi}} \right) |\Pi^{2}|
	 + V_{\text{CF}}|\Phi^{2}| \right)
 \end{equation}
 Taking $\Phi \approx e^{\pm i P_{\phi}\phi}$ we recover the Hamiltonian of 
 the scalar field in flat space-time with the difference of overall factor half.
 \begin{equation}
	 \mathbf{H} =  \frac{1}{2} \left( \frac{\tilde{T}_{\phi}}{2} P_{\phi}^{2}
	 + V_{\text{CF}} \right)
 \end{equation}
 with
 \begin{equation}
	 V_{\text{CF}} = \frac{1}{2} \left( \int d^{3}x \left( |\nabla \tilde{\phi}|^{2} 
	 + m^{2}\tilde{\phi}^{2}	\right)(\vec{x}) \right) \phi^{2}(t) + \int d^{3}x V(\phi)
 \end{equation}
 The Hamiltonian of combined variable theory gives physical 
 time evolution in the flat space-time limit. This is because if we take 
 $\Phi \approx e^{\pm i P_{\phi}\phi} e^{\pm i P^{i}q_{i}}$, combined variable 
 theory reduces to ADM theory. But in absence of 
 gravity, the Hamiltonian gives physical time evolution. \newline
 \textbf{Absence of scalar field:} \newline
 In absence of scalar field $\phi$, the Hamiltonian evolution is a gauge 
 transformation. We take a complex combined variable field.
 \begin{equation}
	 \tilde{T}_{\text{grav}}  F_{ab}\left| \frac{\partial \Phi}{\partial q_{b}} \right|^{2} 
	 + V_{\text{grav}}\left|\Phi\right|^{2} \approx 0
 \end{equation}
 Taking $\Phi \approx e^{\pm i q_{k}P^{k}}$ we recover vector (\ref{eqn21}) as well as scalar 
 Hamiltonian constraints. 
 \begin{equation}
	 \tilde{T}_{\text{grav}}  F_{ab}P^{a}P^{b} + V_{\text{grav}} \approx 0
 \end{equation}
 \textbf{Lapse function, shift vector and combined variable dynamics:}\\
 In order to see the role of lapse function in the Hamiltonian dynamics, find it's 
 conjugate momentum
 \begin{equation}
	 \Pi_{N} \coloneqq \frac{\partial \mathscr{L}}{\partial \left( 
		 \frac{\partial N}{\partial \phi}\right)} = 0
 \end{equation}
 Hamiltonian equations for lapse function and it's conjugate momntum are
 \begin{align}
		& \frac{\partial N}{\partial \phi}  = \lbrace N, \mathbf{H} \rbrace = 0
		& \frac{\partial \Pi_{N}}{\partial \phi} = \lbrace \Pi_{N}, \mathbf{H} \rbrace
 \end{align}
 Since the conjugate momentum of lapse function itself is a primary constraint, the second equation
 gives secondary constraints. This shows that the lapse function does not evolve in the combined variable dynamics. \\
 \hspace{0.5cm} As discussed in (\ref{eqn22}), the shift vector does not play any role in the combined variable 
 dynamics. It puts restrictions on the spatial part of metric $q_{ab}(t, \vec{x})$ and 
 it's conjugate momentum.\\
 \textbf{Discussion: }In ADM theory, 4-dimensional curvature scalar splits into intrinsic 
 curvature and extrinsic curvature parts. In combined variable theory, intrinsic curvature
 $R^{(3)}$ plays a role of potential energy. It is simliar to that of a role of $m^{2}$ in the scalar field 
 theory. Temporal parts of 3-metric $q_{ab}$ along with the scalar field forms a 4-dimensional superspace. 
 Combined variables spread over the space of $q_{a}(t)$ and evolve with $\phi(t)$. The extrinsic curvature
 part decides the distribution of combined variables in this space of $q_{a}(t)$. The scalar field energy 
 plays a role of kinetic energy.
 
 \section{Quantum Theory}
The Hamiltonian of combined variable theory allow us to interpret it as a collection of 
infinitely many quantum harmonic oscillators. This is achieved by defining creation and annihilation
operators with coupling part of the last term replaced with unsettled function 
$\omega(\phi, q_{a})$. The extra quadratic coupling term appears due to the second term 
in the right hand side of (\ref{eqn31}) . These two terms together are the third term in the Hamiltonian function. 
 Therefore, $\omega$ is chosen in such a way that satisfies (\ref{eqn32}). Non-trivial 
 commutation relation between $\Phi$ and $\Pi$ is given as
\begin{equation}\label{eqn27}
	\left[ \Phi(\phi, \vec{q}), \Pi(\phi, \vec{q^{\prime}})\right] = i \delta(\vec{q}, \vec{q^{\prime}})
\end{equation}
Combined variables $\Phi$, $\Pi$, creation and annihilation operators, Hamiltonian are all operators is to 
be understood. Notation $\hat{}$ is avoided for simplicity and introduced in the end.
 Creation and annihilation operators are defined as
\begin{align}\label{eqn28}
	& a \coloneqq \frac{\sqrt{N}}{\sqrt{2}} \left( \sqrt{\frac{2 \sqrt{\text{det } q_{a}}}{\tilde{T}_{\phi}}} \Pi 
	+ i \sqrt{\tilde{T}_{\text{grav}}} \left( \frac{1}{\text{det} q_{a}} \right)^{\frac{1}{4}}  q_{a}\frac{\partial \Phi}{\partial q_{a}} 
	+ i \omega \Phi \right)  \\
	& a^{\dagger} \coloneqq \frac{\sqrt{N}}{\sqrt{2}} \left( \sqrt{\frac{2 \sqrt{\text{det } q_{a}}}{\tilde{T}_{\phi}}} \Pi 
	- i \sqrt{\tilde{T}_{\text{grav}}} \left( \frac{1}{\text{det} q_{a}} \right)^{\frac{1}{4}} q_{a}\frac{\partial \Phi}{\partial q_{a}} 
	- i \omega \Phi \right)
\end{align}
Here $\omega(\phi, q_{a})$ is unsettled function which will be fixed later. 
\begin{align}\label{eqn29}
	a^{\dagger}a =  & \frac{N}{2} \left( \left( \frac{2 \sqrt{\text{det } q_{a}}}{\tilde{T}_{\phi}} \right) \Pi^{2}
	+ 	\tilde{T}_{\text{grav}}  \left( \frac{\partial \Phi}{\partial q_{a}} \right)F_{ab}\left( \frac{\partial \Phi}{\partial q_{b}} \right) 
	+ \omega^{2}\Phi^{2} \right) \\ \nonumber
	&+ \frac{N}{2}  \left( i  \sqrt{2  \frac{\tilde{T}_{\text{grav}}}{\tilde{T}_{\phi}}} 
	\left[ \Pi, q_{a}\frac{\partial \Phi}{\partial q_{a}} \right] 
	+ i\sqrt{ \frac{2}{\tilde{T}_{\phi}}} \left( \text{det }q_{a}\right)^{\frac{1}{4}} \omega
	\left[ \Pi, \Phi \right] \right) \\ \nonumber
	&+ N \left( \frac{1}{\text{det} q_{a}} \right)^{\frac{1}{4}} \sqrt{\tilde{T}_{\text{grav}}} 
	\omega \left( q_{a}\frac{\partial \Phi}{\partial q_{a}}\right)\Phi
\end{align}
Since,
\begin{equation}\label{eqn30}
	\left[ \Pi(\vec{q}), q^{\prime}_{a}\frac{\partial \Phi(\vec{q}^{\prime})}{\partial q^{\prime}_{a}} \right] 
	= D i \delta(\vec{q}, \vec{q}^{\prime})
\end{equation}
$D$ is the number of independent components of 3-metric $q_{ab}$. e.g. for isotropic case
$q_{1}=q_{2}=q_{3}=q$, $D=1$. For $q_{1}\neq q_{2}\neq q_{3}$, $D=3$. 
 Notice that the last term in the (\ref{eqn29}) can be written as 
\begin{equation}\label{eqn31}
	N\left( \frac{1}{\text{det} q_{a}} \right)^{\frac{1}{4}} q_{a} \omega \frac{\partial }{\partial q_{a}}\Phi^{2}  =
	 \frac{\partial }{\partial q_{a}} \left( N \left( \frac{1}{\text{det} q_{a}} \right)^{\frac{1}{4}} q_{a} \omega \Phi^{2} \right) - 
	\left( \frac{\partial }{\partial q_{a}} \left( N \left( \frac{1}{\text{det} q_{a}} \right)^{\frac{1}{4}} q_{a} \omega \right) \right) \Phi^{2}
\end{equation}
When we take integral over the metric space, the first term becomes zero if 
combined variable $\Phi$ is chosen such that $N \left( \frac{1}{\text{det} q_{a}} \right)^{\frac{1}{4}} q_{a} \omega \Phi^{2}$ remains constant
on the surface. Now, first three terms along with the last term in integral
of (\ref{eqn29}) give Hamiltonian if we set 
\begin{equation}\label{eqn32}
	\frac{N}{2} \omega^{2} - \sqrt{\tilde{T}_{\text{grav}}} \left( \frac{\partial }{\partial q_{a}} 
	\left( N \left( \frac{1}{\text{det} q_{a}} \right)^{\frac{1}{4}} q_{a} \omega \right) \right)
	  = \frac{1}{2} N V_{\text{CF}}
\end{equation}
Other two terms in the integral of (\ref{eqn29}) are delta functions and both together is interpreted as a vacuum energy. 
The unsettled $\omega$ is thus a solution to above Riccati equation. 
\begin{equation}\label{eqn33}
	\int  d^{D}q \hspace{0.2cm} a^{\dagger}a = \mathbf{H} 
	 + \int d^{D}q N \left( -D \sqrt{\frac{\tilde{T}_{\text{grav}}}{2\tilde{T}_{\phi}}} 
	+ \omega \left( \text{det }q_{a}\right)^{\frac{1}{4}} \sqrt{ \frac{1}{2\tilde{T}_{\phi}}}  \right) 	\delta(0) 
\end{equation}
Or
\begin{equation}\label{eqn34}
	\mathbf{H}  = \int  d^{D}q \hspace{0.2cm} a^{\dagger}a + 
	 \int d^{D}q N \sqrt{ \frac{1}{2\tilde{T}_{\phi}}} 
	 \left( D \sqrt{\tilde{T}_{\text{grav}}} - \omega \left( \text{det }q_{a}\right)^{\frac{1}{4}} \right) 	\delta(0) 
\end{equation}
The second term $\delta(0)$ corresponds to vaccum energy. 
Only non-trivial commutation relation between creation and annihilation operators is given as 
\begin{equation}\label{eqn35}
	\left[ a(\phi, \vec{q}), a^{\dagger}(\phi, \vec{q^{\prime}}) \right]   = 
	N \sqrt{ \frac{1}{2\tilde{T}_{\phi}}} \left( D \sqrt{\tilde{T}_{\text{grav}}} 
	- \omega  \left( \text{det }q_{a}\right)^{\frac{1}{4}} \right) 
	\delta(\vec{q}, \vec{q^{\prime}})
\end{equation}
A role of creation and annihilation operator changes when 
$ D \sqrt{ \tilde{T}_{\text{grav}}}$ is less than $ \omega  \left( \text{det }q_{a} \right)^{\frac{1}{4}}$.
 In that case, number operator $\hat{n}$ 
is defined as $n \propto aa^{\dagger}$ and for $D \sqrt{\tilde{T}_{\text{grav}}} 
\geq \omega  \left( \text{det }q_{a}\right)^{\frac{1}{4}} $, it is $n \propto a^{\dagger}a$.
 Define number operator $n$ as 
 \begin{align}\label{eqn36}
	 a^{\dagger}a = n \left( N \sqrt{ \frac{1}{2\tilde{T}_{\phi}}} 
	 \left( D \sqrt{\tilde{T}_{\text{grav}}} - 
	 \omega \left( \text{det }q_{a}\right)^{\frac{1}{4}} \right) \right) 
	 & \hspace{0.5cm} \text{for}\hspace{0.25cm} D \sqrt{\tilde{T}_{\text{grav}}} 
	 \geq \omega  \left( \text{det }q_{a}\right)^{\frac{1}{4}} \\ 
	 aa^{\dagger} = n \left( N \sqrt{ \frac{1}{2\tilde{T}_{\phi}}} 
	 \left( \omega \left( \text{det }q_{a}\right)^{\frac{1}{4}} \right) -D \sqrt{\tilde{T}_{\text{grav}}} \right) 
	 & \hspace{0.5cm} \text{for}\hspace{0.25cm} D \sqrt{\tilde{T}_{\text{grav}}} 
	 < \omega  \left( \text{det }q_{a}\right)^{\frac{1}{4}}
 \end{align}
 Then the Hamiltonian without a vaccum energy term is written as
 \begin{equation}\label{eqn37}
	\hat{\mathbf{H}} = \int d^{D}q \left| N \sqrt{ \frac{1}{2\tilde{T}_{\phi}}}
		\left( D \sqrt{\tilde{T}_{\text{grav}}} -
		\omega \left( \text{det }q_{a}\right)^{\frac{1}{4}} \right) \right| \hat{n}
\end{equation}
This Hamiltonian is a collection of infinitely many quanta of the combined variable 
field. The spectrum is geometric. The state of single quantum is represented by 
$| n, q_{a}, \phi \rangle$. $n$ represents quantum number of combined variable field quantum
with metric $q_{a}$ and scalar field $\phi$. Creation operator acting on the vacuum 
produces a single combined variable field quantum.
\begin{equation}
	\hat{a}^{\dagger}_{( q_{a}, \phi)} | 0 >  = | 1,  q_{a}, \phi \rangle
\end{equation}
Annihilation operator acting on the vacuum state gives 0. \\
\hspace*{0.5cm}Adjoint operator of creation operator $\hat{a}$ for real
combined variables is 
\begin{align*}
	 \frac{\sqrt{N}}{\sqrt{2}} \left( \sqrt{\frac{2 \sqrt{\text{det } q_{a}}}{\tilde{T}_{\phi}}} \Pi 
	- i \sqrt{\tilde{T}_{\text{grav}}} 
	\left( \frac{1}{\text{det} q_{a}} \right)^{\frac{1}{4}} q_{a}\frac{\partial \Phi}{\partial q_{a}} 
	- i \omega^{\star} \Phi \right)
\end{align*}
which is not an annihilation operator defined in the (\ref{eqn28}) unless 
$\omega$ is real. Therefore only real solutions to the Riccati equation make 
Hamiltonian a self-adjoint operator. This Riccati equation arises in the process of 
quantization as an inevitable condition. An application of proper boundary conditions
on this Riccati equation guarantees uniqueness of the quantum theory. 
\section{FLRW $\kappa = 0$ cosmology}
Here, the standard ADM theory is re-writtten in terms of temporal part $q(t)$ of 
3-metric and its canonical conjugate momentum. $q(t)$ is related to the scale factor
$a(t)$ by $q = a^{2}$. 
FRW metric is given as
\begin{equation}
	g_{\mu \nu} = \text{diag}( 1, -q(t), -q(t)r^{2}, -q(t)r^{2}\sin(\theta))
\end{equation}
Assume massless scalar field which is constant everywhere. \\
\textbf{ADM theory:} \newline
Choose lapse function $N = 1$ and shift vector $N^{a} = 0$. Total Hamiltonian is then
\begin{equation} \nonumber
	H_{\text{total}}= H_{\phi} + H_{\text{scalar}} 
	= \tilde{T}_{\phi} \frac{P_{\phi}^{2}}{2\sqrt{\text{det }q_{a}}} 
	- \tilde{T}_{\text{grav}} \left( P^{a}\frac{q_{a}q_{b}}{\sqrt{\text{det }q_{a}}}P^{b}\right)
\end{equation} 
\begin{equation} 
	H_{\text{total}} = \frac{P_{\phi}^{2}}{2q^{\frac{3}{2}}} 
	- \frac{1}{3} q^{\frac{1}{2}}P^{2} 
\end{equation}
It is already shown in \ref{item5}, section II, (2.7) and (2.8) that the ADM theory is invariant 
under two rescaling \ref{item5}, section II, (2.6). Therefore omitted fiducial volume part.
Equations of motion are 
\begin{align}
	& \dot{\phi} = \lbrace \phi,  H_{\text{total}} \rbrace =  \frac{P_{\phi}}{q^{\frac{3}{2}}} 
	& \dot{P}_{\phi} = \lbrace P_{\phi},  H_{\text{total}} \rbrace = 0 \\ 
	& \dot{q} = \lbrace q,  H_{\text{total}} \rbrace = -\frac{2}{3} q^{\frac{1}{2}}P 
	& \dot{P} = \lbrace P,  H_{\text{total}} \rbrace =
	\frac{3P_{\phi}^{2}}{4q^{\frac{5}{2}}} 
	- \frac{1}{6}  \frac{P^{2}}{q^{\frac{1}{2}}}
\end{align}
\begin{align}
	\ddot{q} =  - \frac{2}{3} \left( \frac{\dot{q}}{2q^{\frac{1}{2}}}P 
	+q^{\frac{1}{2}}\dot{P} \right)
\end{align}
Plugging equations of $\dot{q}$ and $\dot{P}$ into above equation and using the fact that
$H_{\text{total}}=0 $, we get 
\begin{equation}
	\ddot{q} = -\frac{\dot{q}^{2}}{2q}
\end{equation}
and $H_{\text{total}} = 0$ guarantees $\frac{\dot{q}^{2}}{q^{2}} = \frac{2}{3} \rho_{\phi}$. 
where $\rho_{\phi} \coloneqq \dot{\phi}^{2}$ is scalar field energy density.
Equation for scalar field energy density becomes
\begin{equation}
	\dot{\rho}_{\phi} = -3 \left( \frac{\dot{q}}{q}\right) \rho_{\phi}
\end{equation}
Since the scalar field is massless and constant everywhere, $\rho_{\phi} = p_{\phi}$. 
The standard results of FLRW $\kappa =0$ models are 
 $\frac{\dot{a}^{2}}{a^{2}} = \frac{1}{6}\rho_{\phi}$ (10.73, \ref{item8}),
 $\frac{\ddot{a}}{a} = - \frac{1}{3}\rho_{\phi}$ (10.80, \ref{item8}) with 
$\frac{\ddot{a}}{a} = -2\frac{\dot{a}^{2}}{a^{2}}$ and (10.82, \ref{item8}). 
Results obtained above can be easily verified by taking $q = a^{2}$.\\
\textbf{Discussion: } \\
According to the standard FLRW flat universe model in the ADM form, 
$\dot{q}>0$ tells that the Universe expands in presence of scalar field and 
$\ddot{q}<0$ suggests that this rate of expansion decreases with time. In absence of the 
scalar field, the Universe would have remained static.  
Refer chapter 10 of (\ref{item8}) for further detailed analysis of FLRW models.\\
\textbf{Classical combined variable field:} \newline
Action for the combined variable field defined by (\ref{eqn20}) which has symmetrized gravitational 
part of a field equation is given by
\begin{equation}
	\mathscr{A} = \int d\phi \int dq \hspace{1mm} \frac{1}{2} 
	\left( \frac{\nu_{0}}{2q^{\frac{3}{2}}} \left( \frac{\partial \Phi}{\partial \phi} \right)^{2}
	-  \frac{1}{3} \nu_{0} q^{\frac{1}{2}} \left( \frac{\partial \Phi}{\partial q} \right)^{2} 
	 \right)
\end{equation}
$D = 1$ because gravitational part of superspace is 1 dimensional. 
It gives following field equation
\begin{equation}
	\left( \frac{\nu_{0}}{2 q^{\frac{3}{2}}} \frac{\partial^{2}}{\partial \phi^{2}} 
	-   \frac{1}{3} \nu_{0} q^{\frac{1}{2}}	\frac{\partial^{2}}{\partial q^{2}} 
	-  \frac{1}{6} \nu_{0} q^{-\frac{1}{2}} \frac{\partial}{\partial q} 
	 \right) \Phi (\phi, q) = 0 
\end{equation}
The superspace $(\phi, q)$ is 2 dimensional. 
In the limit $q\rightarrow 0$, the first term is dominant and $\Phi \rightarrow A\phi + B$. 
In the limit $q\rightarrow \infty$, the other two terms are dominant and 
$\Phi \rightarrow Aq^{\frac{5}{9}} + B$. Energy density and pressure for this combined variable 
field are given as 
\begin{equation} \label{eqnrhocos}
	\rho = \frac{\nu_{0}}{4q^{\frac{3}{2}}} \left( \frac{\partial \Phi}{\partial \phi} \right)^{2}
	+  \frac{1}{6} \nu_{0} q^{\frac{1}{2}} \left( \frac{\partial \Phi}{\partial q} \right)^{2}
\end{equation}
\begin{equation}
	p = \frac{\nu_{0}}{4q^{\frac{3}{2}}} \left( \frac{\partial \Phi}{\partial \phi} \right)^{2}
	- \frac{1}{6} \nu_{0} q^{\frac{1}{2}} \left( \frac{\partial \Phi}{\partial q} \right)^{2}
\end{equation}
Energy density $\rho$ as well as pressure has singularity, the `Big Bang singularity' at $q = 0$.
The Big Bang singularity exists even in absence of the scalar field because 
$\Phi = Aq^{\frac{5}{9}} + B$. Therefore the second term diverges at the origin.
 In the limit $q \rightarrow \infty$, $\rho \rightarrow 0$ and $p \rightarrow 0$ with 
$p = -\rho$. This is the equation of state for pure gravitational field described by 
FLRW $\kappa =0$ metric in absence of the scalar field. \\
\textbf{Discussion: }\\
The combined variable field $\Phi$ behaves more like a scalar field in early development of 
the Universe. As Universe expands the scalar field energy density decreases 
and gravitational field energy density starts dominating. In absence of the scalar field,
momentum (\ref{eqn23}) conjugate to combined variable field becomes primary constraint and  
$\frac{\partial \Phi}{\partial \phi} = 0$. Evolution is not a physical evolution. There is 
nothing with respect to which gravitational field can evolve and situation becomes static.
In other words, the gravitational field evolves resulting into the expansion of the Universe. 
Later, scalar field energy density tends to zero. In absence of scalar field,
gravitational field cannot evolve. \\
\textbf{Quantum combined variable field:} \newline
Quantization of the theory is straight forward once Riccati equation (\ref{eqn32})
 is solved. $\omega$ which
 is the solution to the Riccati equation allows us to write the Hamiltonian in terms of creation
and annihilation operators. 
\begin{equation}
	\frac{1}{2}\omega^{2} - \sqrt{ \frac{1}{3}\nu_{0}}
	\frac{\partial}{\partial q}\left( q^{\frac{1}{4}}\omega \right) = 0
\end{equation}
 Solution to this equation is given as
\begin{equation}
	\omega = - \sqrt{ \frac{1}{3}\nu_{0}}
	\frac{1}{q^{1/4} \left( \sqrt{q}- c_{1}\right)}
\end{equation}
The Hamiltoninan for FRW $\kappa = 0$ is
\begin{equation}
	\hat{\mathbf{H}}  = 
		\int dq \left| \sqrt{ \frac{1}{6}} \left( 1 +
		\frac{1}{\sqrt{q}-c_{1}}
		\right) \right| \hat{n}
\end{equation}
The Hamiltonian is a collection of infinitely many combined variable field 
quantum. A value of constant $ c_{1}$ can be fixed by applying an initial condition to the 
Riccati equation. The energy is quantized in the units of 
\begin{equation}
	h =  \sqrt{ \frac{1}{6}} \left( 1 + \frac{1}{\sqrt{q}- c_{1}}\right)
\end{equation}

\textbf{Discussion: } Classically, $q |_{t=0} \coloneqq c_{1}^{2} = 0$ at the time of Big Bang. Therefore, the energy spectrum of the quatum has singularity at the Big bang. But the Universe is a collection of these quantum, the total energy of the Universe is collection of all these quantum. Therefore, the minimum $q$ the Universe can have depends on total energy of the Universe. In other words, the Universe must have began at $q\neq 0$. 
\section{Conclusion}
\hspace*{0.5cm}If the Klein-Gordon field is a result of quantizing relativistic particles
then, the combined variable field is a result of quantizing gravitational field  
together with the scalar field. Unlike ADM theory where field equations tell us 
how gravitional field and scalar field evolve relative to one another, the combined 
combined variable theory tells, both gravitational and scalar field combined to form 
a combined variable field which is distributed over $q_{a}(t)$ and evolve relative to $\phi(t)$. 
On one particle ($\Phi \approx e^{\pm i\phi P_{\phi}}e^{\pm iq_{a} P^{a}}$) like 
interpretation, the combined variable theory reduces to the ADM theory.
Hamiltonians are different for observers in two different frames and are
related to each other through lapse function. Shift vectors does not play 
any role in the combined variable dynamics. It puts restrictions on the spatial
part of gravitational field variables. \newline
\hspace*{0.5cm} In the absence of scalar field $\phi$, Hamiltonian constraints get recovered
and combined variable field becomes static. In the absence of 
gravity, combined variable field behaves as a standard (special relativistic) scalar 
field and Hamiltonian gives physical time evolution. \newline
\hspace*{0.5cm} The theory applied to FLRW $\kappa =0$ model in order to understand 
it's implications. The classical combined variable theory agrees with the standard ADM theory.
Both classical theories are different viewpoint of the same reality. The quantum of the combined variable
theory is a quantum of both scalar and field and gravitational field together. For $q \rightarrow \infty$,
it can be interpreted as a quantum of gravity.
\section{acknowledgement}
I am immensely grateful to Prof. Ajay Patwardhan for providing expertise. This work would not have possible 
without his guidance.

\end{document}